\documentclass[12pt]{article}
\usepackage{amssymb,amsmath}
\newtheorem{theorem}{Theorem}
\begin{document}
\title{Locally covariant quantum field theory
\thanks{to appear in the Proceedings of the International Conference on 
Mathematical Physics, Lisbon 2003}}
\author{Klaus Fredenhagen\\
{\small II. Institut f\"ur Theoretische Physik, D-22761 Hamburg, 
klaus.fredenhagen@desy.de}}  

\date{}
\maketitle
\abstract{The principle of local covariance which was recently 
introduced admits a generally covariant formulation of quantum field theory.
It allows a discussion of structural properties of quantum field theory 
as well as the perturbative construction of renormalized interacting models 
on generic curved backgrounds and opens in principle the way 
towards a background independent perturbative quantization of gravity.}

\section{Introduction}

Quantum field theory may be understood as the incorporation 
of the principle of locality (``Nahwirkungsprinzip''), which 
is at the basis of classical field theory, into quantum 
theory, formally encoded in the association of points $x$ of 
spacetime to observables $\varphi(x)$ of quantum theory \cite{Haag}. The 
physical picture behind is that experiments in a laboratory 
are analyzed in terms of spacetime positions of measuring 
devices, as for instance in scattering theory. 

This picture assumes an a priori notion of spacetime. Such 
a notion is in conflict with principles of general 
relativity where a point should be characterized by 
intrinsic properties of geometry, in the spirit of Leibniz, 
and in contrast to Newton's idea of an absolute spacetime. 
In order to incorporate gravity into quantum theory one 
therefore should aim at a formulation of the theory which 
uses only intrinsic and local properties of spacetime.

The traditional formalism of quantum field theory, however, 
is full of nonlocal concepts which have no obvious local 
geometric interpretation. Most remarkably, the concepts of 
a vacuum and of particles are meaningful only for 
spacetimes which have special structures. As a consequence, 
an S-matrix cannot be introduced, in the generic case. On the more 
technical side, the euclidean formulation of quantum field 
theory and the corresponding path integral admit a 
covariant formalism on compact Riemannian spaces, but, in 
general, there is no analogue of the 
Osterwalder-Schrader-Theorem \cite{OS} which allows the transition to 
a spacetime with a Lorentzian metric.

There seem to be two reasons for the use of nonlocal 
concepts in quantum field theory. One is the occurence of 
divergences which prohibit a direct translation of the 
intrinsically local classical field theory into quantum 
theory; the other are the nonlocal features of quantum 
physics, as they become visible in the violation of Bell's 
inequalities.

The problem of singularities can be solved by replacing 
momentum space techniques by methods of microlocal analysis. 
This program was successfully carried through by Brunetti 
and Fredenhagen \cite{BF} on the basis of a crucial 
observation of Radzikowski \cite{Radzikowski}. The 
nonlocal aspects of quantum physics can be circumvented by 
shifting the emphasis from the states (with the essentially 
nonlocal phenomenon of entanglement) to the algebras of 
observables where locality is encoded in algebraic 
relations. We therefore developed a generalization of the 
framework of algebraic quantum field theory \cite{BFV,HW} 
which is 
\begin{itemize}
        \item  intrinsically local

        \item  generally covariant

        \item  based on Haag's concept of algebras of local 
        observables
\end{itemize}
We show that the concept can be realized in renormalized 
perturbation theory, and that structural theorems of 
quantum field theory as the spin statistics connection 
\cite{V} and  
the existence of PCT symmetry \cite{H} can be formulated and, 
in a somewhat weaker form, proven. Moreover, we indicate 
how the theory can be physically interpreted. Some tentative 
remarks on a consistent framework for perturbative 
quantization of gravity conclude the paper.   

\section{Quantum field theory as a functor}

Let me first recall the formalism of algebraic quantum field 
theory (``local quantum physics''). There the basic 
structure consists in an association of spacetime 
regions $\mathcal{O}\subset\mathcal{M}$ to algebras of 
observables $\mathfrak{A}(\mathcal{O})\subset \mathfrak{A}(\mathcal{M})$ 
satisfying the Haag-Kastler axioms \cite{HK,D}:
\begin{displaymath}
        \mathcal{O}_{1}\subset \mathcal{O}_{2} \Longrightarrow 
        \mathfrak{A}(\mathcal{O}_{1}) \subset \mathfrak{A}(\mathcal{O}_{2}) 
        \text{  (isotony)}
\end{displaymath}
\begin{displaymath}
        \mathcal{O}_{1} \text{ spacelike to } \mathcal{O}_{2} 
        \Longrightarrow [\mathfrak{A}(\mathcal{O}_{1}) ,
\mathfrak{A}(\mathcal{O}_{2}) ] = \{0\} \text{ (locality)}
\end{displaymath}

The identity component of the group $G$ of isometries $g$ of 
$\mathcal{M}$ is represented by automorphisms $\alpha_{g}$ of 
$\mathfrak{A}(\mathcal{M})$ such that
\begin{displaymath}
        \alpha_{g}(\mathfrak{A}(\mathcal{O})) \subset 
        \mathfrak{A}(g\mathcal{O})
\end{displaymath}

If $\mathcal{O}$ is a neighbourhood of a Cauchy surface of 
$\mathcal{M}$ then
\begin{displaymath}
        \mathfrak{A}(\mathcal{O}) = \mathfrak{A}(\mathcal{M})  \text{ (time 
        slice axiom)} 
\end{displaymath}

A map $\mathfrak{A}$ with the properties above is called a 
local net. It is the principle of algebraic quantum field 
theory that the local net characterizes the theory uniquely. 
This principle has been tested mainly on Minkowski space and 
was found to be satisfied in all cases considered so far 
(see e.g. the contribution of Buchholz to the proceedings of 
ICMP2000 in London \cite{Bu}). 

In the application of the framework of algebraic quantum 
field theory to curved spacetime, there is, however, the 
problem that the group $G$ of isometries is trivial in the 
generic case, and that its replacement by the group of 
diffeomorphisms is in conflict with the axiom of locality.

We therefore proposed in \cite{BFV} the following 
generalization of the Araki-Haag-Kastler framework of 
algebraic quantum field theory (see also \cite{V} and 
\cite{HW}): Instead of defining the theory on a specific 
spacetime, we define it on all spacetimes (of a suitable 
class) in a coherent way. Let $\mathfrak{Man}$ denote 
a category whose objects are globally hyperbolic spacetimes 
which are oriented and time oriented and whose arrows are 
isometric embeddings of one spacetime into another one, such 
that orientations and causal relations are preserved. 

A locally covariant quantum field theory is now defined as a 
covariant functor from $\mathfrak{Man}$ into a category 
$\mathfrak{Alg}$ of operator algebras, where the arrows are 
faithful homomorphisms. It depends on the problem under 
consideration whether one assumes the objects to be 
C*-algebras or topological algebras.

This definition contains the original Haag-Kastler framework. 
Namely, if one restricts the 
functor to the full subcategory whose objects are 
globally hyperbolic, causally convex subsets of a fixed 
spacetime $\mathcal{M}$, one obtains a local net which 
satisfies the conditions of isotony and covariance. 

The analogue of the remaining Haag-Kastler axioms can 
easily be formulated. The axiom of locality (in the sense of 
commutativity of spacelike separated observables) means 
that, provided $\psi_{i}:\mathcal{M}_{i}\to\mathcal{N}$, $i=1,2$ are 
arrows of $\mathfrak{Man}$ such that the images of 
$\mathcal{M}_{1}$ and $\mathcal{M}_{2}$ are spacelike separated in 
$\mathcal{N}$, the corresponding subalgebras of 
$\mathfrak{A}(\mathcal{N})$ commute. The time slice axiom, 
on the other hand, says 
that if the image of $\psi$ is a neighbourhood of a Cauchy 
surface of the target spacetime, then the corresponding 
subalgebra already coincides with the full algebra.

The latter property allows a comparison of time evolution on 
different spacetimes. Namely, let $\mathcal{M}_{1}$ and 
$\mathcal{M}_{2}$ have Cauchy surfaces with mutually isometric 
neighbourhoods $\mathcal{N}_{\pm}$, and let 
$\psi_{i,\pm}:\mathcal{N}_{\pm}\to\mathcal{M}_{i}$ denote the 
corresponding embeddings. Then, 
using the fact that the corresponding embeddings 
$\mathfrak{A}\psi_{i,\pm}=:\alpha_{i,\pm}$ of operator 
algebras are surjective, we can define an automorphism of 
$\mathfrak{A}(\mathcal{M}_{1})$ by
\begin{displaymath}
        \beta= \alpha_{1,+}\alpha_{2,+}^{-1} 
        \alpha_{2,-} \alpha_{1,-}^{-1} \ .
\end{displaymath}
$\beta$ describes the effect of the change of the metric 
from $\mathcal{M}_{1}$ to $\mathcal{M}_{2}$ 
between the two Cauchy surfaces and is called the relative Cauchy evolution. 
It may be interpreted as a 
time evolution in the interaction picture. In particular, 
if $\beta_{h}$ arises from the addition of a symmetric 
tensor $h$ with compact support to the original metric, 
then its functional derivative at $h=0$ can be interpreted as a 
commutator with the energy momentum tensor ( up to a factor 
of $\frac{1}{2i}$). In fact, it is always covariantly 
conserved, and for the free Klein-Gordon field it coincides 
with the adjoint action of the canonical energy momentum 
tensor ($\cdot\frac{1}{2i}$) \cite{BFV}. 

Based on the locally covariant system of operator algebras 
described by the functor $\mathfrak{A}$, one can introduce 
other locally covariant structures. They are, by 
definition, natural transformations between functors 
defined on $\mathfrak{Man}$. 

First of all, two locally covariant theories are 
equivalent, if the functors are naturally equivalent. Namely, 
let $\mathfrak{A}_{1}$ and $\mathfrak{A}_{2}$ be two functors 
from $\mathfrak{Man}$ to ${\frak Alg}$. They are naturally 
equivalent, if there exists a family $\gamma_{\mathcal{M}}$ of 
isomorphisms from $\mathfrak{A}_{1}(\mathcal{M})$ to 
$\mathfrak{A}_{2}(\mathcal{M})$
such that
\begin{displaymath}
        \gamma_{\mathcal{M}}\circ\mathfrak{A}_{1}\psi = {\frak 
        A}_{2}\psi \circ \gamma_{\mathcal{M}}\ .
\end{displaymath}

One may also define a locally covariant state to be a 
family $\omega_{\mathcal{M}}$ of states on  
$\mathfrak{A}(\mathcal{M})$ with the covariance property
\begin{displaymath}
                \omega_{\mathcal{M}}=\omega_{\mathcal{N}}\circ 
                \mathfrak{A}\psi \ .
\end{displaymath}
But it turns out, that states with this property do not 
exist in typical cases. This may be interpreted as 
nonexistence of a vacuum state in a generally covariant 
framework, and is the root of many problems of quantum field 
theories on curved spacetimes; in particular, a covariant path 
integral, even in a formal perturbative sense, does not exist. 

One can however show, for the free Klein Gordon field, 
that a locally covariant minimal folium of states does exist. 
A folium of states on a C*-algebra $\mathfrak{B}$ is a norm closed set of 
states which is invariant under the operations     
\begin{displaymath}
   \omega\to\omega_{B}=
   \frac{\omega(B^{*}\cdot B)}{\omega(B^{*}B)}\ ,
   \ B\in\mathfrak{B} \ .               
\end{displaymath} 
and under convex combinations. A folium is minimal, if it 
does not contain proper subfolia. We have the following 
theorem \cite{BFV}, which might be considered as a version of the 
principle of local definiteness \cite{HNS}.
\begin{theorem}
  For the Klein-Gordon theory, there exists a family of 
  folia $\mathfrak{F}_{\mathcal{M}}$ 
  of $\mathfrak{A}(\mathcal{M})$ such that
  \begin{displaymath}
        (\mathfrak{A}\psi)^{*}{\frak F}_{\mathcal{N}} \subset 
        \mathfrak{F}_{\mathcal{M}}
  \end{displaymath}
  and is minimal if $\psi(\mathcal{M})$ is relatively compact.
\end{theorem}           
   
   The most important natural transformations are the local 
  quantum fields. A locally covariant scalar field $A$ is a 
  family of linear continuous maps $A_{\mathcal{M}}:\mathcal{D}(\mathcal 
  {M})\to \mathfrak{A}(\mathcal{M})$ (here we prefer to work with 
  topological operator algebras), such that
  \begin{displaymath}
        \mathfrak{A}{\psi}\circ A_{\mathcal{M}}= A_{\mathcal{N}}\circ 
        \mathcal{D}\psi
  \end{displaymath}
  where $\mathcal{D}\psi\equiv\psi_{*}$ denotes the push forward on the 
  test function spaces,
  \begin{displaymath}
        (\psi_{*}f)(x)=\left\{
        \begin{array}{ccc}
                f(\psi^{-1}(x)) & , & x\in\psi(\mathcal{M})  \\
                0 & , & \mathrm{else} \ .
        \end{array}\right.
  \end{displaymath}
  Vector, tensor and spinor fields are analogously defined 
  as natural transformations with the functor of spaces of 
  compactly supported sections of suitable bundles. One may 
  also look at nonlinear functionals, an 
  example being the relative Cauchy evolution $\beta$. 
  There one introduces, for every spacetime $\mathcal{M}$, the 
  set  $\mathfrak{Lor}(\mathcal{M})$ of Lorentz metrics which differ 
  from the given metric 
  only on a compact region. Let $\beta_{\mathcal{M},g}$ be the 
  relative Cauchy evolution between $\mathcal{M}$ and the 
  spacetime obtained by replacing the metric by $g\in \mathfrak 
  {Lor}(\mathcal{ M})$, and let 
  $\beta_{\mathcal{ M}}:\mathfrak{ Lor}(\mathcal{ M})
  \to \mathrm{Aut}(\mathfrak{ A}(\mathcal{ M}))$ be defined by 
  \begin{displaymath}
        \beta_{\mathcal{M}}(g) =\beta_{\mathcal {M},g} \ .
  \end{displaymath}     
  Then $\beta$ is a natural transformation in the sense that
  \begin{displaymath}
        \mathfrak{A}\psi\circ \beta_{\mathcal{M},g}=
        \beta_{\mathcal{N},\psi_{*}g}\circ\mathfrak{A}\psi \ .
  \end{displaymath}
 %%%%%%%%%%%%%%%%%%%%%%%%%%%%%%%%%%%%%%%%%%%%%%%%%%%%%%%%%
  \section{Free fields and Wick polynomials}
  As a first example we construct the theory of the free 
  Klein Gordon field. There the algebra 
  $\mathfrak{A}(\mathcal{M})$ is the unital *-algebra generated 
  by elements $\varphi_{\mathcal{M}}(f)$, 
  $f\in \mathcal{D}({\mathcal{M}})$ with the following relations:
  First we require that the map
  \begin{displaymath}
        f\to\varphi_{\mathcal{M}}(f) 
   \end{displaymath}
   is linear and the involution is given by
   \begin{displaymath}
        \varphi_{\mathcal{M}}(f)^{*} =
        \varphi_{\mathcal{M}}(\overline{f}) \ .
   \end{displaymath}
   Then we assume that the field equation is 
   satisfied in the weak sense,
   \begin{displaymath}
        \varphi_{\mathcal{M}}(Kf)=0
    \end{displaymath}
    with the Klein Gordon operator $K$ on $\mathcal{M}$.  
    Finally, the  
    commutation relations are given by
    \begin{displaymath}
        [\varphi_{\mathcal{M}}(f),\varphi_{\mathcal{M}}(g)] = 
        i(f,E_{\mathcal{M}}g)
    \end{displaymath}
    where $E_{\mathcal{M}}$ is the fundamental solution of the 
    Klein-Gordon equation on $\mathcal{M}$. 
    
    The action $\mathfrak{A}\psi\equiv\alpha_{\psi}$ of $\mathfrak{A}$ on 
   arrows $\psi:\mathcal{M}\to \mathcal{N}$ is given
    by
    \begin{displaymath}
        \alpha_{\psi}(\varphi_{\mathcal{M}}(f))= \varphi_{\mathcal 
        {N}}(\psi_{*}f) \ .
     \end{displaymath}
     The crucial fact which guarantees that $\alpha_{\psi}$ 
     is an homomorphism of algebras, is that the 
     restriction of the fundamental solution $E_{\mathcal{N}}$ 
     to $\psi(\mathcal{M})$ coincides with the push forward of 
     $E_{\mathcal{M}}$. 
     
     By this construction, we also obtained a locally 
     covariant field, namely 
     $\varphi=(\varphi_{\mathcal{M}})_{\mathcal{M}}$. 
     
     One may now ask whether there exists a locally 
     covariant state which then could be chosen as the 
     analogoue of the vacuum state. In order to exclude 
     pathologies we restrict ourselves to states 
     $\omega_{\mathcal{M}}$ whose 
     $n$-point functions $\omega_{\mathcal{M}}^{(n)}$ are 
     distributions on $\mathcal{M}^n$,
     \begin{displaymath}
        \omega_{\mathcal{M}}(\varphi_{\mathcal{M}}(f_{1})\cdots 
        \varphi_{\mathcal{M}}(f_{n})) =
        \langle \omega_{\mathcal{M}}^{(n)},f_{1}\otimes 
        \cdots \otimes f_{n}\rangle \ .
     \end{displaymath}  
     The 2-point function on Minkowski space $\mathbb{M}$ is then, due to 
     the required covariance property, Poincar\'{e} invariant and must 
     therefore be the 
     usual vacuum 2-point function 
     $\omega_2^{\mathbb{M}}(x,y)=\Delta_{+}(x-y)$. On the 
     cylinder spacetime $\mathrm{Cyl}=\mathbb{R}\times \mathbb{T}^3$, 
     on the other 
     hand, covariance requires that the 2-point function is invariant 
     under space and time translations,
     \begin{displaymath}
       \omega_2^{\mathrm{Cyl}}(x,y)=G(x-y) \ ,
     \end{displaymath}
     where $G$ is a distribution on Minkowski space which is 
     periodic with period $2\pi$ in all spatial coodinates.     
     If $G$ coincided on subregions 
     which are isometrical to subregions of Minkowski space with $\Delta_+$ 
     it had to be in the time variable the boundary value of an 
     analytic function on the upper halfplane, hence it must be a 
     ground state 2 point function of the cylinder spacetime, 
     which in the massive case is unique and is given by
     \begin{displaymath}
         \Delta_{+}^{\mathrm{Cyl}}(x)=
         \sum_{n\in\mathbb{Z}^3}\Delta_+(x+2\pi(0,n))
     \end{displaymath} 
     and thus does not coincide with $\Delta_+$ in an open region. In the 
     massless case a ground state on the cylinder spacetime does not exist 
     because of zero modes.

     We conclude that no locally covariant Hadamard state of the free 
     Klein-Gordon field exists. This proves the nonexistence of 
     a vacuum state on generic curved spacetimes mentioned in the 
     previous section.. 
         
     Also the familiar concept of a particle has no 
     obvious locally covariant formulation. Therefore the standard 
     interpretation of field theory in terms of scattering of particles
     does not apply to generic curved spacetimes. We may use, however, 
     the existence of other locally covariant fields.

     In the case of the Klein-Gordon field, one may try to define in 
     addition to the free field itself other members of the Borchers class.
     In \cite{BFK} Wick polynomials of the free field were defined as 
     operator valued distributions. Their definition depended on the 
     choice of a specific Hadamard state. In the case of the Wick square 
     the definition was 
     \begin{displaymath}
       \varphi^2_{\omega}(x)=
       \lim_{y\to x}\left(\varphi(x)\varphi(y)-\omega_2(x,y)\right) \ .
     \end{displaymath}
     Two such Wick squares with respect to different Hadamard states 
     $\omega,\omega'$ differ by a smooth function $H_{\omega,\omega'}$
     which satisfies the covariance condition
     \begin{displaymath}
       H_{\omega\circ\alpha_{\psi},\omega'\circ\alpha_{\psi}}(x)=
       H_{\omega,\omega'}(\psi(x))
     \end{displaymath}
     and the cocycle condition
     \begin{displaymath}
       H_{\omega,\omega'}+H_{\omega',\omega''}+H_{\omega'',\omega}=0 \ .
     \end{displaymath}
     The problem is now to find functions $h_{\omega}$ 
     which transform covariantly
     \begin{displaymath}
       h_{\omega\circ\alpha_{\psi}}(x)=h_{\omega}(\psi(x))
     \end{displaymath}
     and trivialize the cocycle condition,
     \begin{displaymath}
       h_{\omega}-h_{\omega'}=H_{\omega,\omega'} \ .
     \end{displaymath}
     Then the definition
     \begin{displaymath}
       \varphi^2=\varphi^2_{\omega}-h_{\omega}
     \end{displaymath}
     gives a locally covariant field.

     A solution was given by Hollands and Wald \cite{HW} in terms of the 
     construction of Hadamard states. A Hadamard state has a 2-point 
     function of the form 
     \begin{displaymath}
       \omega_2 = \frac{u}{\sigma} +v \ln\sigma +w
     \end{displaymath}
     with smooth functions $u,v,w$ where $u$ and $v$ are determined by 
     local geometry and 
     $\sigma$ is the square of the geodesic distance between the two 
     arguments. One then can choose 
     \begin{displaymath}
       h_{\omega}(x)=w(x,x) \ .
     \end{displaymath}
%%%%%%%%%%%%%%%%%%%%%%%%%%%%%%%%%%%%%%%%%%%%%%%%%%%%%%%%%%%%%%%%%%%
\section{Renormalized perturbation theory}
Renormalized perturbation theory can be formulated as the 
construction of the local S-matrix as a formal power series of time ordered 
products
\begin{displaymath}
  S(g\mathcal{L})= \sum\frac{i^n}{n!}\int dx\, g(x_1)\cdots g(x_n)\, 
  T\mathcal{L}(x_1)\cdots \mathcal{L}(x_n) 
\end{displaymath}
where $\mathcal{L}$ is the Lagrangian  and $g$ is a test function with 
compact support. The time ordered products are operator valued distributions 
which are well defined on noncoinciding points. In this approach 
which is due to St\"uckelberg \cite{S}, 
Bogoliubov \cite{B}, Epstein and Glaser \cite{EG}, the ultraviolet 
divergences show up  as ambiguities in the extension to 
coinciding points. The local S-matrices can then be used for a 
construction of local algebras of observables, see e.g. \cite{DF}. 

The construction can be generalized to curved 
spacetimes by methods of microlocal analysis \cite{BF}. In the case of 
Minkowski space the ambiguities are restricted by the requirement of 
Poincar\'{e} invariance; as shown by Hollands and Wald \cite{HW} a 
corresponding 
restriction in the general covariant situation can be obtained by 
imposing the condition of local covariance. 

Namely, if $\psi$ is a morphism from $\mathcal{M}$ to $\mathcal{N}$, then
one requires
\begin{displaymath}
  \alpha_{\psi}(S(g\mathcal{L}_{\mathcal{M}}))= 
   S(\psi_{*}g \, \mathcal{L}_{\mathcal{N}}) \ .
\end{displaymath}
This condition fixes the ambiguities up to terms which depend locally on the 
metric. By additional assumptions concerning the scaling behaviour, 
continuity and analyticity, Hollands and Wald were able to show that the 
renormalization freedom on generic curved spacetimes is up to a 
possible coupling to curvature terms the same as in Minkowski space.
%%%%%%%%%%%%%%%%%%%%%%%%%%%%%%%%%%%%%%%%%%%%%%%%%%%%%%%%%%%%%%%%%%%%
\section{Spin-statistics, PCT, and all that}
The principle of local covariance also admits a formulation of important 
structural theorems of quantum field theory \cite{SW}. An example is the 
connection between spin and statistics. The proof on Minkowski space 
heavily relies on the structure of the Poincar\'{e} group, so at first sight 
a generalization to curved spacetimes seems to be hopeless. But actually 
Verch \cite{V} was able to use the covariance principle to relate 
the statistics on a generic spacetime to that on Minkoski spacetime and 
thus obtained an elegant proof of a general spin statistics theorem.

Another famous theorem of general quantum field theory is the PCT theorem. 
Since it involves elements of the Poincar\'{e} group even the formulation 
of the theorem on generic curved spacetimes is unclear. But in the 
locally covariant framework one may associate to every time oriented spacetime
$\mathcal{M}\in\mathrm{Obj}(\mathfrak{Man})$ the manifold 
$\overline{\mathcal{M}}$ by reversing the time orientation. One then defines 
the PCT transformed functor $\overline{\mathfrak{A}}$ by
\begin{displaymath}
  \overline{\mathfrak{A}}(\mathcal{M})=
  \overline{\mathfrak{A}(\overline{\mathcal{M}})}
\end{displaymath}
where for an algebra over the complex numbers the overlining means 
complex conjugation.

The statement of PCT invariance would then be that the functors 
$\mathfrak{A}$ and $\overline{\mathfrak{A}}$ are naturally equivalent.

A first result in this direction was obtained by Hollands \cite{H}. He 
could show that at least the operator product expansions 
(provided they exist) in both theories are equivalent.

%%%%%%%%%%%%%%%%%%%%%%%%%%%%%%%%%%%%%%%%%%%%%%%%%%%%%%%%%%%
\section{Outlook}
We have seen that the standard interpretation of quantum field theory 
in terms of scattering of particles is not meaningful on generic spacetimes.
Instead one should interprete the theory in terms of locally covariant fields.
As an example one may use the Buchholz-Ojima-Roos theory of local 
equilibrium states \cite{BOR}. One selects a subspace $V$ of locally covariant 
fields and compares a given state $\omega$ on a given spacetime with 
convex combinations of homogeneous KMS states on Minkowski space. 
If at a point $x$ the expectation values of 
fields $A\in V$ can be described by a probability measure on the set 
of homogenous KMS states $\sigma$,
\begin{displaymath}
  \omega(A_{\mathcal{M}}(x)) = \int d\mu(\sigma)\sigma(A_{\mathbb{M}}(0)) 
   \ , A\in V \ ,
\end{displaymath}
this measure can be interpreted in terms of local thermodynamic properties 
of $\omega$ at the point $x$, as measured by $V$. In particular this 
provides a definition of temperature in nonstationary spacetimes. 

The proposed framework also allows the formulation of the action of 
compactly supported diffeomorphisms. 
Let $\beta_{\mathcal{M},g}$ be the relative Cauchy evolution induced 
by a change of the metric of $\mathcal{M}$ as defined in Sect.3.
If $g$ arises from the application of a 
compactly supported diffeomorphism $\phi$, $g=\phi_{*}g_{\mathcal{M}}$, 
then by the general covariance of the formalism we find
\begin{displaymath}
  \beta_{\mathcal{M},\phi_{*}g_{\mathcal{M}}}=\mathrm{id}
\end{displaymath}
One may also introduce quantities which transform covariantly under 
diffeomorphisms. An example are the retarded fields which were 
introduced by Hollands and Wald \cite{HW}.
They are given by
\begin{displaymath}
  A_{\mathcal{M},g}^{\mathrm{ret}}(f)=
  \alpha_{\psi}\alpha_{\psi_g}^{-1} (A_{\mathcal{M}_g}(f)) \ .
\end{displaymath}
where $\psi,\psi_g$ denote appropriate embeddings of another spacetime onto
a neighbourhood of a Cauchy surface which lies sufficiently far in the 
past such that $g$ differs from $g_{\mathcal{M}}$ only in its future. If 
now the metric is changed by a compactly supported diffeomorphism $\phi$, 
then the retarded field transforms as
\begin{displaymath}
   A_{\mathcal{M},\phi_{*}g_{\mathcal{M}}}^{\mathrm{ret}}(f)
   = A_{\mathcal{M}}(\phi^{*}f) \ .
\end{displaymath}

One may ask whether the covariant framework which was presented here 
also allows a generally covariant perturbative approach to the 
quantization of gravity. This in fact possible. Namely, one has to 
quantize the theory on all spacetimes simultaneously. The fact that the 
metric is dynamical means that the split into background metric and 
quantized fluctuations is arbitrary. This can be expressed by the postulate 
that the relative Cauchy evolution $\beta_g$ is the identity for all 
metrics $g\in\mathfrak{Lor}(\mathcal{M})$. A corresponding consideration 
applies to other background fields, see e.g. \cite{M}.

\section*{Acknowledgments}

The present contribution relies on joint work with Romeo Brunetti and 
Rainer Verch and on an intense exchange of ideas with Stefan Hollands and 
Robert Wald.


\begin{thebibliography}{99}


\bibitem{Haag} R. Haag: Local quantum physics. Springer 1996 (2. Edition)
\bibitem{OS} K. Osterwalder, R. Schrader, 
        {\it Commun. Math. Phys. }{\bf 31}, 83 (1973).
\bibitem{BF} R. Brunetti, K. Fredenhagen, 
        {\it Commun. Math. Phys. }{\bf 208}, 623 (2000).
\bibitem{Radzikowski} M. Radzikowski, 
        {\it Commun. Math. Phys. }{\bf 179}, 529 (1996).
\bibitem{BFV} R. Brunetti, K. Fredenhagen, R. Verch, 
         {\it Commun. Math. Phys. }{\bf 237}, 31 (2003).
\bibitem{HW} S. Hollands, R. Wald, 
        {\it Commun. Math. Phys. }{\bf 223}, 289 (2001),
        {\it Commun. Math. Phys. }{\bf 231}, 309 (2002).
\bibitem{V} R. Verch, {\it Commun. Math. Phys. }{\bf 223}, 261 (2001).
\bibitem{H} S. Hollands, {\it Commun. Math. Phys. }{\bf 244}, 209 (2004).
\bibitem{HK} R. Haag, D. Kastler, {\it J. Math. Phys. }{\bf 5}, 848 (1964).
\bibitem{D} J. Dimock, {\it Commun. Math. Phys. }{\bf 77}, 219 (1980).
\bibitem{Bu} D. Buchholz, math-ph/0011044
\bibitem{HNS} R. Haag, H. Narnhofer, U. Stein, 
        {\it Commun. Math. Phys. }{\bf 94}, 219 (1984).
\bibitem{BFK} R. Brunetti, K. Fredenhagen, M. K\"ohler, 
        {\it Commun. Math. Phys. }{\bf 180}, 633 (1996).
\bibitem{S} E. C. G. St\"uckelberg, A. Petermann,
        {\it Helv. Phys. Acta }{\bf 26}, 499 (1953).
\bibitem{B} N. N. Bogoliubov, D. Shirkov: Introduction to the Theory 
         of Quantized Fields, New York: John Wiley and Sons, 
         1976, 3rd edition 
\bibitem{EG} H. Epstein, V. Glaser, 
        {\it Annales Poincar\'{e} }{\bf A19}, 211 (1973).
\bibitem{DF} M. D\"utsch, K. Fredenhagen, 
        {\it Commun. Math. Phys. }{\bf 219}, 5 (2001).
\bibitem{SW} R. F. Streater, A. Wightman, PCT, Spin and statistics 
        and all that, New York: Benjamin 1964
\bibitem{BOR} D. Buchholz, I. Ojima, H. Roos, 
        {\it Annals of Physics }{\bf 297}, 219 (2002).
\bibitem{M} P. Marecki, hep-th/0312304 

\end{thebibliography}
\end{document}